\documentclass[11pt,a4paper]{article}
\pdfoutput=1 
\usepackage{jinstpub}
\usepackage{graphicx}
\usepackage{lineno}

\title{Synthesis of radio signals from extensive air showers using previously computed microscopic simulations}

\author[1,2,3]{Matias Tueros}
\author[3]{Anne Zilles}

\affiliation[1]{IFLP - CCT La Plata - CONICET, Casilla de Correo 727 (1900) La Plata, Argentina}
\affiliation[2]{Depto. de Fisica, Fac. de Cs. Ex., Universidad Nacional de La Plata,  Casilla de Coreo 67 (1900) La Plata, Argentina}
\affiliation[3]{ Institut d'Astrophysique de Paris, CNRS-Sorbonne Universit\'e, 98 bis boulevard Arago, F-75014 Paris, France }

\emailAdd{tueros@fisica.unlp.edu.ar}
\abstract{
The detection of extensive air showers (EAS) through their radio signal is becoming one of the most promising techniques for the study of Neutrinos and Cosmic rays at the highest energies.
For the design, optimization and characterization of radio arrays, and of their associated reconstruction algorithms, tens of thousands of Monte Carlo simulations  are needed. 
Current available simulation codes can take several days to compute the signals produced by a single shower, making it impossible to produce the required simulations in a reasonable amount of time, in a cost-effective and environmental-conscious way.
In this article we present a method to synthesize the expected signals (the full time trace, not just the peak amplitude) at any point around the shower core, given a set of signals simulated in a finite number of antennas strategically located in a pattern that exploits the signature features of the radio wavefront. The method can be applied indistinctly to the electric field or to the antenna response to the electric field, in the three polarization directions.
The synthesized signal can be used to evaluate trigger conditions, compute the fluence or reconstruct the shower incoming direction, allowing for the production of one single library of simulations that can be used and re-used for the characterization and optimization of radio arrays and their associated reconstruction methods, for a thousandth part of the otherwise required CPU time.}


\begin{document}
\maketitle
\flushbottom

\section{Introduction}
\label{Sec:Intro}

When a high-energy cosmic particle interacts with the molecules in the atmosphere, an extensive shower of secondary particles is generated. Moving close to the speed of light, the charged particles in the shower are deflected by the Earth's magnetic field, collectively producing coherent electromagnetic pulses that can be detected at large distances from the shower in the VHF (30-300 MHz) frequency range with a relatively simple and cheap radio detector. This makes the radio detection of cosmic particles a very promising technique for instrumenting the large surface areas required by the next generation of cosmic rays and neutrino detectors \cite{TimsReview,SchroderReview}.

To assess the capabilities of a given detector array, the signals expected to be produced by the particle showers needs to be computed covering all the possible primary particle energies, the possible incoming directions and the possible particle types (from now on, the incoming particle phase-space) with enough statistics to account for the variability from shower-to-shower fluctuations, requiring thus the simulation of several tens of thousands of particle showers. 

Furthermore, the optimization of the detection capabilities of a given array can require changing the separation between detectors, the array total size and general layout, the possibility to include in-fills or voids in the array, the location of the detectors on the terrain, etc. A change in any of these parameters (from now on, the detector phase-space) will impact the performance of reconstruction algorithms, a factor that also needs to be studied and taken into account to make decisions on the final detector design. Covering the detector phase-space requires re-evaluating the signals produced by the showers in each different configuration.

This problem is common to all EAS detection experiments, regardless of their detection technique. However, the simulation of radio signals presents an additional challenge. In particle detectors for example, the generated signal depends only on the particles arriving at the detector. This means that if the type, position, energy and arrival direction of the particles reaching the surface are stored, the output of the shower simulation can then be re-used to compute the signal at any position, in any particle detector.

For radio detectors, the electric field produced at a given location is the sum of the contributions of \textit{all} the charged particles in the shower, and depends on the relative position of the detector to the trajectories of the particles. This means that to separate the simulation of the shower from the simulation of the signals it produces, the trajectories of all the charged particles needs to be stored. To preserve the coherence information of the radio emission, the trajectories need to be detailed to better than the characteristic wavelength, and their timing to better than fraction of their period. Even using statistical thinning methods to keep the number of simulated particles at bay, this means tracking charge, position, time and energy for typically 10$^6$ simultaneous trajectories along 10$^5$ m. This would require, for a single shower, storage space several TBytes, which is at the limit of current technology and totally impractical when libraries of several tens of thousand of showers are required.

One way to circumvent this problem is to use a "macroscopic" approach. In this approach, the spatial and temporal distributions of the charged particles in the shower, or a parameterization of these, is used to solve the electro-dynamical equations and compute the resulting electric field at each detector location. This removes the need for storage of all the particle trajectories, replacing them by histograms or a relatively small set of parameterizations that tries to describe the overall characteristics of the particle cascade \cite{Scholten2008,Scholten2018,Marin2012,deVries2013}.

This simplification makes macroscopic approaches very fast, and very useful to gain insights in the processes responsible for the features of the radio signal. However, the assumptions used to simplify the problem and represent the individual particles with charge distributions erase the fine temporal and spatial structures in the signal, and require the fine-tuning of free parameters that affect the results and limit their predictive power \cite{Scholten2008}.

If the limitations of macroscopic approaches are to be avoided, and the storage of the particle trajectories is impractical, the simulation of the radio signals needs to be done in parallel with the simulation of the particle shower. Consequently, each time the location of the detectors is changed, the simulation needs to be repeated.

The simulation of the electric field following the contribution of each individual particle (the "microscopic" approach \cite{ZHAireS,CoREAS}) is CPU intensive. The computation of the electric field produced at a single position takes nearly as much CPU time as the computation of the whole particle shower itself. Since the computations at each antenna position are almost independent of each other, the required CPU time scales with the number of desired positions. The simulation of a single event involving a few dozens of antennas can easily take one CPU-day. Repeating the simulation of thousands of events to cover the detector phase-space can easily require millennia of CPU time. While achievable in large supercomputing centers, when available, this is costly and requires lots of CO$_2$-emitting electric power and hardware. Looking for a way to re-utilize microscopic simulations, while keeping the accuracy of its results, is thus paramount.

Attempts at re-utilizing the results from microscopic simulations to synthesize the radio signal already exist, in different stages of development. One method \cite{Butler2017} computes separately the signals at a given detector position originating from different slices of the longitudinal development of the shower, so that the total signal is the sum of the signals from all the slices. Next, the spectral features of the signal in each slice are fitted to produce a signal model that can be applied to synthesize the signals of a shower with a different longitudinal profile but with the same geometry, at the same antenna position. While promising, this method is still not suited for its general use, as the fit of the model parameters require human supervision on a case by case basis. If the fit procedure were to be automated, and the method proves to produce very accurate synthesized signals, an interpolation strategy similar to the one proposed later in this article could be used to get the signal at different antenna positions.

Another attempt at re-utilizing the results from microscopic simulations is the method known as radio-morphing \cite{RadioMorph}. This method tries not only to re-utilize the simulations, but also to drastically reduce the number of simulations needed to cover the incoming primary phase-space by re-scaling or "morphing" the signals simulated on a reference shower to the signals expected for a different shower. The morphing is achieved by using the known variation of the signal amplitude with primary energy, geomagnetic angle and air density, as well as taking into account the variation in the angular opening of the Cherenkov cone. Simulating the signals in a wisely chosen set of positions, the signal at any point in space can then be synthesized by interpolation, allowing the re-use of the reference simulation as needed. The method was successfully used for the synthesis of signals from showers initiated by neutrinos at the GRAND array \cite{WhitePaper}, with a mean bias in the estimation of the peak amplitude of the signals of less than 10\% and a standard deviation of 25\% (albeit some significant outliers), which is impressive considering that a single shower was used to cover all the incoming phase-space.
 
Radio-morphing is bound to provide better results if more showers are used as reference, reducing the amount of "morphing" that needs to be done for going from the reference shower to a desired shower. In the limit, the error from "morphing" the shower can be completely removed by simulating a reference shower for every desired incoming configuration. This would negate completely the method's capability to reduce the number of simulations required to cover the incoming primary phase-space, but would keep the ability to re-use the simulations to synthesize the signals required to study the detector phase-space with improved accuracy.

The method presented in this paper is an implementation of this extreme approach, optimized for the synthesis of signals on the ground by a more efficient choice of simulated antenna positions and an improved phase interpolation algorithm, with the important addition of an accurate prediction of the synthesized signal arrival time, which was unavailable in the radio-morphing method and is crucial for the study of the performance of geometry reconstruction algorithms.

In this article we will present the basic aspects of radio emission from extensive air showers and their role in the motivation of the developed interpolation strategy, along with the interpolation procedure, in section \ref{Sec:Method}. A general description of the performance of the method using the cross-correlation of the synthesized and simulated signals will be presented in section \ref{Sec:PerformanceCorrelation}. The obtained precision on several signal observables relevant to triggering and reconstruction algorithms, both on a flat surface and for complex topographies, will be shown in section \ref{Sec:Performance}. We finish by giving some concluding remarks in the last section.

\section{Underlying model and strategy for the interpolation}
\label{Sec:Method}

The radio emission of particle showers can be described as a superposition of two distinct emerging phenomena. The most important one, accounting for 80 - 95 \% of the emission, is the appearance of a charge dipole that changes with time as the positive and negative charges in the shower are deflected in opposite directions by the Earth's magnetic field $\vec{B}$, the so-called Geomagnetic effect \cite{Geomagnetic,TimsReview}.

The second effect, known as the Askaryan effect \cite{Askaryan1,Askaryan2}, is the build-up of a net negative charge in the shower front as electrons ripped-off from the molecules in the air are incorporated into the shower front, while positrons annihilate with electrons in the atmosphere. 

The Geomagnetic emission is linearly polarized in the direction perpendicular to the direction of $\vec{B}$ and the shower direction $\vec{v}$. Since the dipole increases as more and more charge is separated and then decreases as the particles loose their energy and are removed from the shower, the resulting emission is bipolar. The Askaryan emission, which is akin to a current pulse on the shower axis, is also linearly polarized but in this case the polarization vectors are radially oriented perpendicular to the shower direction.

The emission region, the shower front, is only some meters thick and produces coherent radiation up to several MHz. The particles in the shower front travel almost at the speed of light in vacuum $c$,  giving rise to a time compression of the received signal in the direction of the Cherenkov angle $\alpha_{c}$ (roughly 1 degree in air), pushing the signal bandwidth up to the GHz range around that particular direction \cite{deVries2011,Alvarez2012,Nelles2015}.

The emission pattern of particle showers can be modeled, as a first approximation, as the emission of a point like source moving at $c$ along the direction of $\vec{v}$, with an amplitude that is proportional to the number of particles in the shower \cite{WashingtonTM}. This simple model implies that the bulk of the emission originates close to $X_{max}$, the position where the shower reaches its maximum development. 

Considering a constant index of refraction $n$, and ignoring the contribution from the Askaryan effect, the amplitude of the radiation pattern in a plane perpendicular to $\vec{v}$ will be rotationally symmetric around the shower core. The radiation will be polarized in the direction of $\vec{v}\times\vec{B}$, and propagating as a spherical wavefront at the speed of light in the medium, $v=c/n$. 

In this simplified model, due to the rotational symmetry, the signal at any point in a plane perpendicular to the propagation direction of the shower can be described solely as a function of $\alpha$, the angle between the point and the shower axis, measured from $X_{max}$ (Fig. \ref{Fig:Geometry}).  We will express $\alpha$ in units of the Cherenkov angle so that $\alpha_{c}=1$, regardless of the value of n.

In a plane oblique to the propagation direction of the shower, such as the ground plane, the rotational symmetry is broken due to the difference in distance to the emission point for points at the same $\alpha$. Consequently, to describe the signal in the ground plane its necessary to introduce a second coordinate, the polar angle $\phi$, whose origin is chosen along the projection of $\vec{v}\times(\vec{v}\times\vec{B})$ on the ground. The amplitude pattern varies more abruptly in  $\alpha$, in particular close to $\alpha_{c}$, while the variations in $\phi$ are smoother. 

We have ignored so far the Askarian emission, as it contributes on average with less than 10\% \cite{Schellart2014} of the total emission. The amplitude of the Askarian emission is radially symmetric, but its polarization is pointing towards the shower axis, and will add constructively to the Geomagnetic emission for some values of $\phi$ and destructively for others, further disrupting the rotational symmetry. 

\subsection{Interpolation Strategy}
\label{sec:strategy}

The interpolation strategy that we will use to synthesize the signal $S_i(\alpha_i,\phi_i)$ at a point on the ground plane with coordinates ($\alpha_i,\phi_i$) is to choose four signals, $S_I(\alpha_I,\phi_I)$ , $S_{II}(\alpha_{II},\phi_{II})$, $S_{III}(\alpha_{III},\phi_{III})$ and $S_{IV}(\alpha_{IV},\phi_{IV})$, such that

\begin{figure}
  \includegraphics[width=15cm]{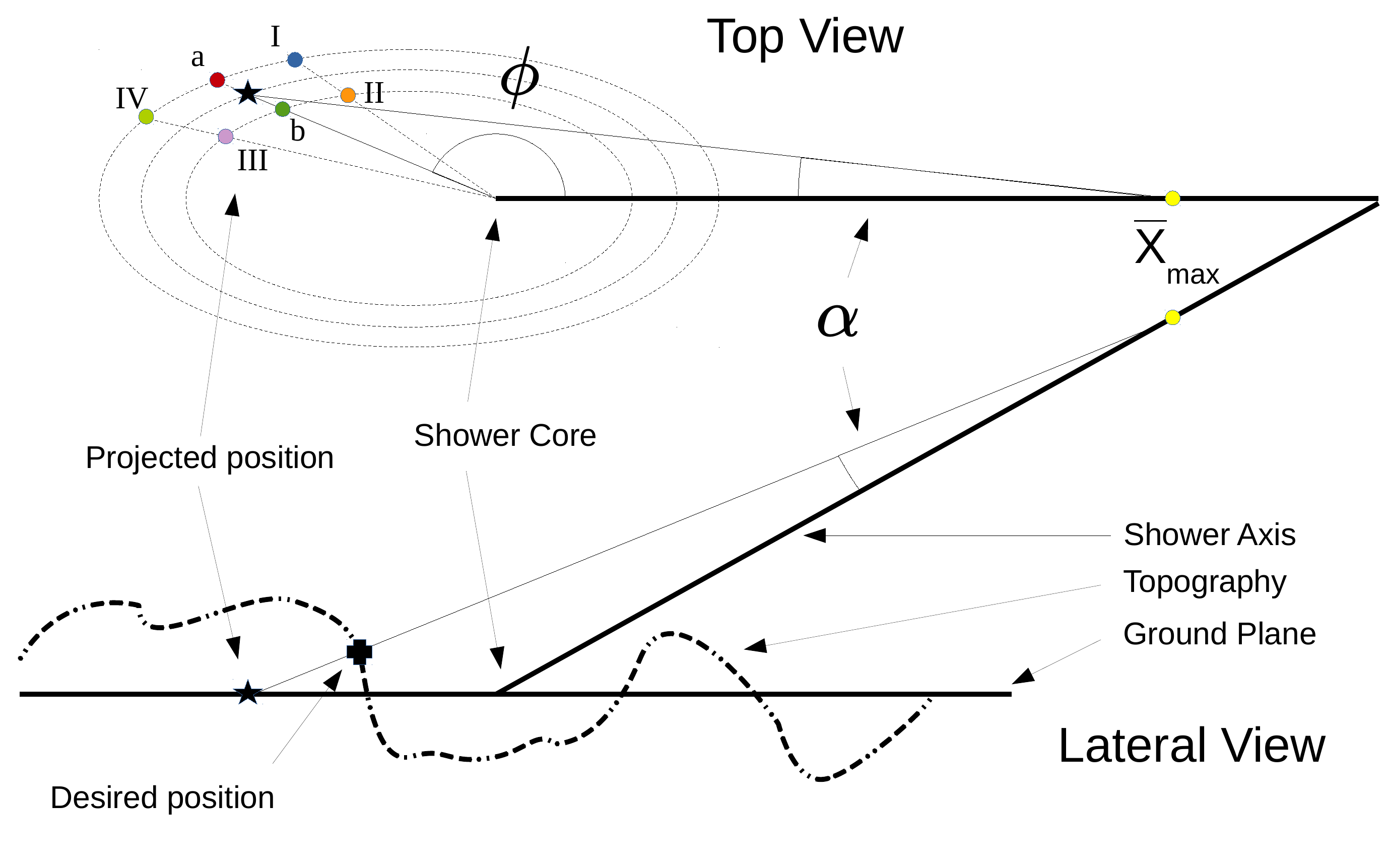}
  \caption{ Shower geometry and interpolation strategy. To get the signal at a given position (cross) on the terrain, the signal at the projection of this position into the ground plane (star) is synthesized. This is done using the signals at points I and IV to get the signal at \textbf{a} and the signals at II and III to get signal at \textbf{b}. The signals at \textbf{a} and \textbf{b} are then interpolated to get the synthesized signal (star), whose amplitude is then scaled to take into account its true distance to ${X}_{max}$, and get the signal at the desired position (cross)}
  \label{Fig:Geometry}
\end{figure}

\begin{enumerate}

\item $\alpha_I > \alpha_i > \alpha_{II}$ 
\item $\phi_I = \phi_{II} > \phi_i $
\item $\alpha_{IV} > \alpha_i > \alpha_{III} $
\item $\phi_{IV} = \phi_{III} < \phi_i $
\item $\alpha_I = \alpha_{IV}$
\item $\alpha_{II} = \alpha_{III}$
\end{enumerate}

and interpolate first along the $\phi$ coordinate between $S_I$ and $S_{IV}$ to get $S_a(\alpha_I,\phi_i)$ and between $S_{II}$ and $S_{III}$ to get $S_b(\alpha_{II},\phi_i)$, and finally interpolate along the $\alpha$ coordinate between $S_a$ and $S_b$ to get $S_i(\alpha_i,\phi_i)$ (see Fig. \ref{Fig:Geometry}). When using the interpolation procedure described in \ref{Sec:Interpolation}, this  is equivalent to a bilinear interpolation \cite{Bilinear} of the signals in I,II,III and IV in ($\alpha$,$\phi$).

To be able to synthesize the signal at any desired position on the ground, the signal  needs to be simulated in an array of positions in $(\alpha, \phi)$, covering the full range of $\phi$ and the relevant range in $\alpha$ up to a given $\alpha_{max}$. Since the variation of the signal is faster in $\alpha$, this variable needs to be sampled more densely, specially close to the Cherenkov angle. In this work we use a quadratic spacing in $\alpha$ to sample more densely the region where $\alpha<1$. 

\begin{figure}
  \includegraphics[width=15.2cm,trim=0.5cm 0.5cm 1cm 0.5cm, clip]{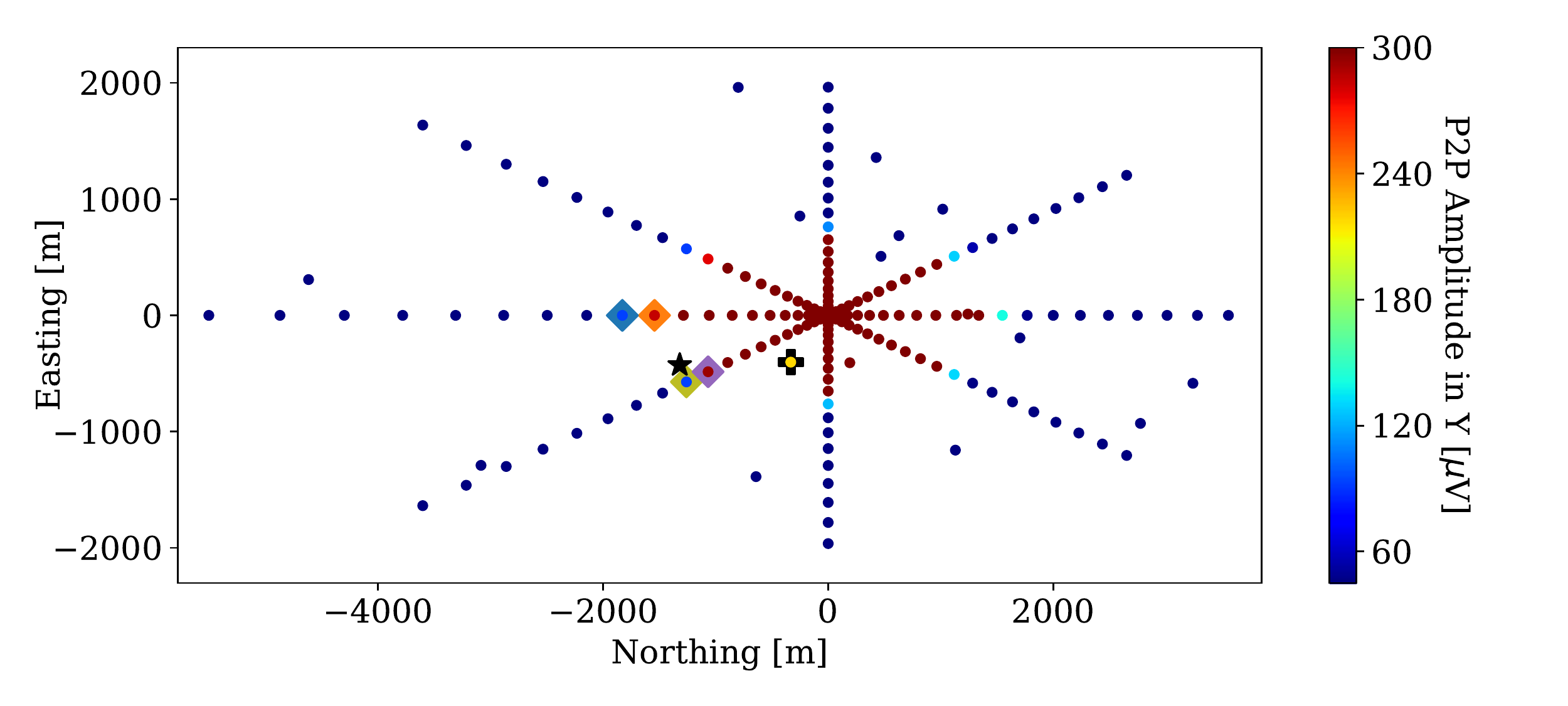}
  \caption{Example conical star-shape plus 16 random test locations from an 4 EeV proton cascade at 63$ ^{\circ}$ zenith incoming from the North, plotted on equal axis scales to emphasize the elliptical shape. For the synthesis of the signal at a given position (cross), which in this case is 450m above the ground plane, the projection to the ground plane is computed (star) and the signal is obtained from the signals in the highlighted postions, as shown in fig \ref{Fig:ExampleInterpolation}}
  \label{Fig:ConicalStarshape}
\end{figure}

A denser array of sampled positions gives smaller interpolation errors, but requires more CPU time to perform the signal simulations, reducing the gains from synthesizing the signals. A good compromise was found to be the use of 8 equally spaced points to sample $\phi$, and 20 points to sample $\alpha$, as shown in fig. \ref{Fig:ConicalStarshape}.

When using 8 values of $\phi$, the resulting pattern on the ground resembles the "star-shape" array frequently used in the characterization of the shower front, but projected on the ground plane along a cone with it's vertex in  $X_{max}$. We will refer to this array as the "conical star-shape", to differentiate it from the more frequently used parallel projection of the star-shape along the shower axis, which does not preserve $\alpha$.

When index of refraction gradients are introduced in the description, the spherical wavefront is distorted by the differences in optical path to each observer position, and the rotational symmetry is further disrupted. The determination of $\alpha_{c}$  is no longer straightforward, as the optical path depends on $\phi$.  We use $\alpha_{c}=\textrm{arcos(}n_{X_{max}}\textrm{)}$, where $n_{X_{max}}$ is the average value of $n$ along the shower axis from $X_{max}$ to the core position on the ground. In the case of an index of refraction varying exponentially with altitude, as is the one used in our simulations, we found that this prescription slightly over-estimates the value of $\alpha_{c}$, ensuring that the Cherenkov peak remains inside the better sampled area of $\alpha<1$, while retaining the scaling benefits of using a unit for $\alpha$ that changes with the position of $X_{max}$.

Since $\alpha_{c}$ is slightly different for each value of $\phi$, the interpolation using a constant $\alpha$ will be less accurate. This will be one of the sources of systematic errors in the method, and could be reduced by a denser sampling of the $\phi$ coordinate at the expense of more CPU time, or using the correct value of $\alpha_{c}$ on each branch of the star-shape, at the expense of more complexity in the method. 

The optimal value of $\alpha_{max}$ depends on the objectives of the study to be done with the synthesized signals. Signal amplitudes at large values of $\alpha$ are usually very small and/or out of the bandwidth of the detectors, except for high energy showers with the maximum very close to the ground. A parameterization of $\alpha_{max}$ with the shower energy and the distance to $X_{max}$ is recommended to avoid wasting CPU time sampling regions of the shower front that will later be discarded by quality cuts.

When taking into account the local topography over a given site or when looking for a suitable site for the detector array along a given geographic region, the signals might need to be synthesized at different altitudes above sea level. To accommodate for this, the star-shape pattern is located on a ground plane at a conveniently chosen intermediate altitude (as shown in Fig. \ref{Fig:Geometry}). When the signal at a location with a different altitude is desired, the signal is synthesized at the projection of the desired position into this ground plane. To compensate for the fact that the synthesized signal is being computed closer or farther away from $X_{max}$ than the desired position, the signal amplitude is scaled with the ratio of the distance to $X_{max}$ of the desired and synthesized positions.

\subsection{Signal Interpolation}
\label{Sec:Interpolation}

The interpolation of radio signals relies on the fact that the electric field produced by a particle shower (or the response of an antenna to it) changes smoothly and continuously between nearby positions. We will use an interpolation method in Fourier space based on the one presented in \cite{EvaTesis} and later used in \cite{RadioMorph}, with a different treatment of the phase interpolation and introducing time alignment of the signals to minimize phase differences and to provide the correct time of arrival.

A signal of length $T$ sampled regularly at intervals $T/N_s$ can be written as a Fourier series

\begin{equation}
    S_j(t) = \frac{1}{N_s} \sum_{k=0}^{N_s-1} s_j(k) e^{2 i \pi k t/T},
\end{equation}

with $N_s$ the number of time samples and the complex Fourier components 

\begin{equation}
s_{j}(k)=|s_{j}(k)|e^{i\theta_j(k)}.
\end{equation}

Given the signal at two reference points $S_1$ and $S_2$ the signal at a desired position between them $S_i$ has interpolated Fourier components 

\begin{equation}
|s_i(k)|=\frac{d_{1}}{d_{1}+d_{2}}|s_{1}(k)|+\frac{d_{2}}{d_{1}+d_{2}}|s_{2}(k)|,
\end{equation}

\begin{equation}
\theta_i(k)=\frac{d_{1}}{d_{1}+d_{2}}\theta_1(k)+\frac{d_{2}}{d_{1}+d_{2}}\theta_2(k),
\end{equation}

where $d_{1}$ and $d_{2}$ are the distances from the desired position to the reference points.

Since $\theta$ is cyclic in $[0,2\pi)$, there is an indetermination in the absolute phase difference of the Fourier components of the signals being interpolated. To reduce the effects of this indetermination, the times of the reference signals are measured relative to the expected time of arrival of the signal at the reference position. 

\begin{figure}
  \includegraphics[width=15.2cm,trim=0 0 0 0, clip]{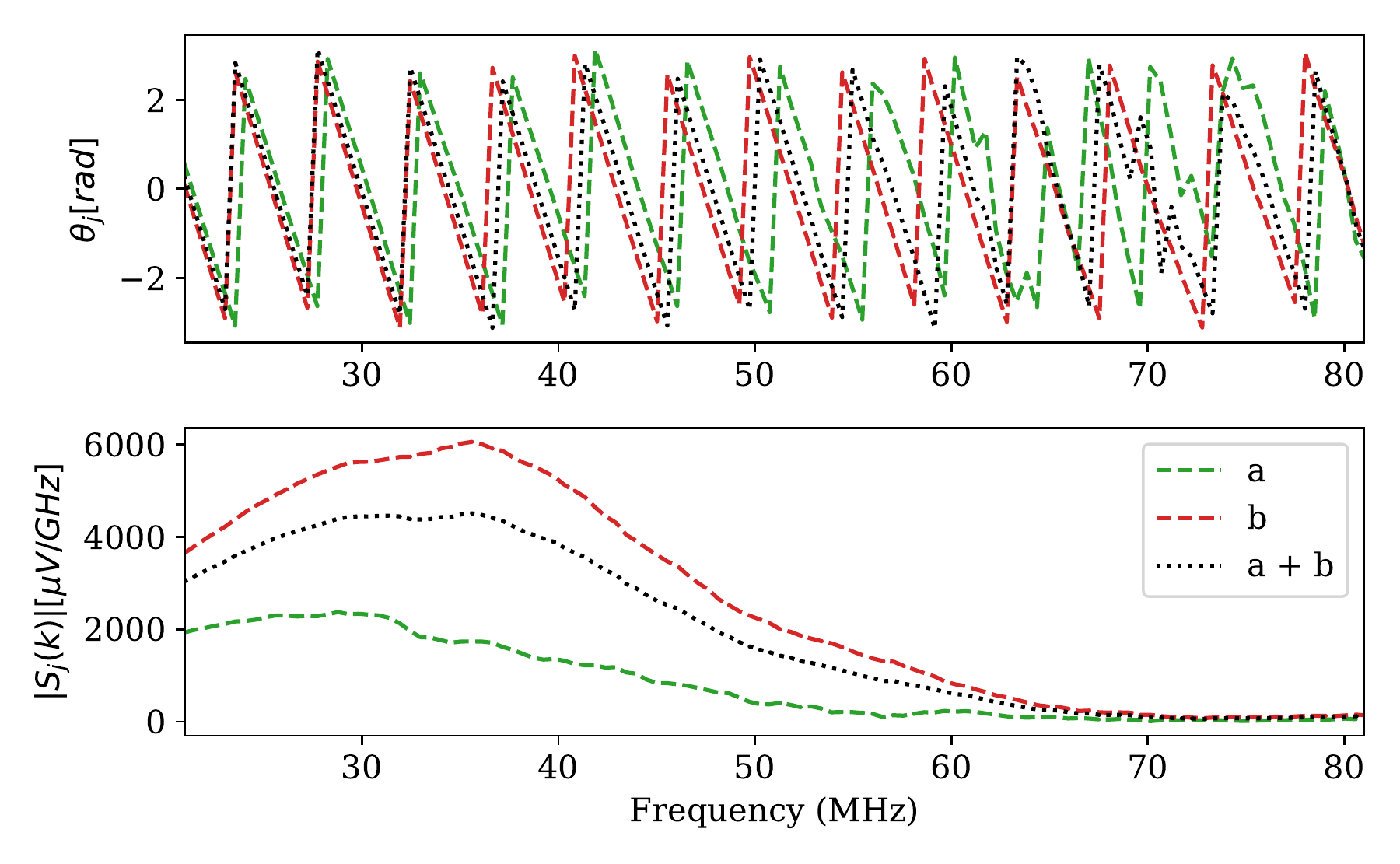}
  \caption{Example of the interpolation in Fourier space of signals in \textbf{a} and \textbf{b} (see fig. \ref{Fig:AmplitudeAlpha}), for the selected antenna of fig. \ref{Fig:ConicalStarshape}. The signals in time domain can be seen in the bottom left of fig. \ref{Fig:ExampleInterpolation}.}
  \label{Fig:ExampleFourier}
\end{figure}

The phase interpolation is performed minimizing the phase difference at each $k$, adding $2\pi$ to $\theta_1$ or $\theta_2$ so that $|\theta_1 - \theta_2|<\pi$, and then warping the resulting phase in the $[0,2\pi)$ range. An example result from this procedure can be seen in fig. \ref{Fig:ExampleFourier} for the for the selected antenna of fig. \ref{Fig:ConicalStarshape}. An example instance of the $\theta$ shifting procedure can be seen at 59MHz, where the phase of \textbf{b} was shifted by $-2\pi$ before the interpolation. This prevented an unnecessary jump in the interpolated phase.
 
According to our simplified model, the emission of the shower originates from a point source at $X_{max}$. Considering a non-dispersive medium and straight line propagation, the expected arrival time at a given location is $n_{eff}.d/c$ , where  $n_{eff}$ is the average value of the index of refraction along the path between $X_{max}$ and the given location and $d$ is the distance along this path. To recover the absolute time at the interpolated position, the interpolated signal is referred to the time of arrival at the interpolated position, using the corresponding $n_{eff}$. Deviations from this simple model, that might make the signal start a little before or after, will automatically be included in the reference signals, and propagated to the interpolation.  

\begin{figure}
  \includegraphics[width=16cm,trim=1.5cm 1cm 0.5cm 1cm, clip]{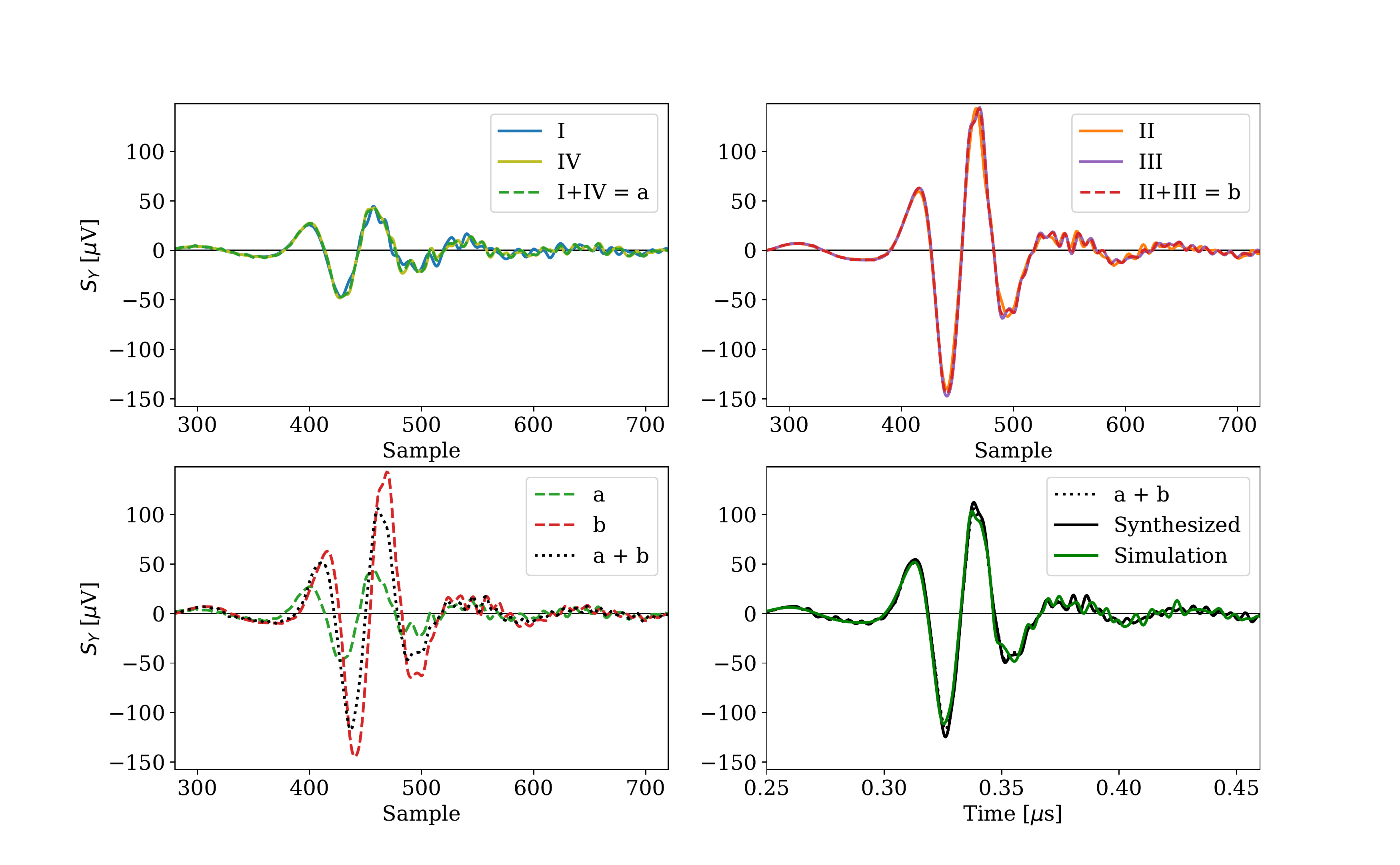}
  \caption{Example of the signal synthesis procedure. Signals in neighboring antennas I,II,III,IV (see fig. \ref{Fig:Geometry}), for the example event of fig. \ref{Fig:ConicalStarshape}, are interpolated into signals \textbf{a} and \textbf{b}, which are then interpolated, time aligned and scaled for the distance to get the synthesized signal. The true signal simulated at the desired antenna position is shown for comparison. The difference between the simulated and the synthesized signal is shown in the center of fig. \ref{Fig:CorrelationExample}.}
  \label{Fig:ExampleInterpolation}
\end{figure}

An example of the interpolation procedure, for the Y (East-West) polarization of the antenna marked with a cross in fig. \ref{Fig:ConicalStarshape} is shown in fig. \ref{Fig:ExampleInterpolation}. Note how similar signals I,IV (and its interpolation a) and signals II,III (and its interpolation b) are, showing the smoothness of the variations in $\phi$. The resulting a and b are more dissimilar, showing the faster signal variation with $\alpha$.

\section{Performance of the method}
\label{Sec:PerformanceCorrelation}

\subsection{Simulations used}
\label{Sec:Library}

To characterize the performance of the method, we produced a library of 3456 showers  using ZHAireS 1.0.28 \cite{ZHAireS1028}, with statistical thinning starting at $E/E_{primary}=10^{-5}$ and the Sybill 2.3c \cite{Sybill23} hadronic interactions model. The showers covered the energy range between $10^{16.3}$eV and $10^{18.6}$eV in 24 logarithmic steps of size 0.1, a 25\% energy increase per step. The zenith angle $\theta$ ranged from 38.23 to 87.1 deg in 16 steps in $\rm{log_{10}(1/cos(\theta))}$, giving each step approximately a 20\% increase in footprint size. Gammas, Protons and Iron nuclei incoming from the North, South and East were used as primary particles. The ground level was set at 1086m above sea level. The refractivity was modeled as an exponential with a value of 325 at sea level and a scale height of 8km. 

For each shower, a conical star-shape of 160 antennas (8 arms, 20 antennas per arm) plus an additional 16 randomly distributed test antennas at ground level were generated. For each position, the simulated electric field was used to compute the response of the HorizonAntenna \cite{WhitePaper} to get the signal in each antenna channel, corresponding to each polarization direction: $S_X$ (North-South), $S_Y$ (East-West) and $S_Z$ (Up-Down). 

Each conical star-shape had its aperture angle ($\alpha_{max}$) adjusted so that  the peak amplitude of the the voltage signal falls below 45$\mu V$ (3 times the expected galactic noise, $\sigma_{noise}$) around antenna number 10-12 of the arm, to guarantee good footprint coverage even if aggressive trigger scenarios are being studied.

At each test antenna the signal was synthesized following the prescription presented in this work, giving a total of 55296 synthesized signals in each channel to be compared with the simulation.

\subsection{Zero Normalized Cross Correlation}
To assess the similarity of the synthesized ($S_{i}$) and simulated ($S_{s}$) signals, we will study their Zero Normalized Cross-Correlation r($\tau$), defined for the zero-normalized signals $S^{'}_{j}=S_{j}-\mu_{j}$ as

\begin{equation}
    r(\tau)=\int{ \! \frac{(S^{'}_{s}(t))(S^{'}_{i}(t-\tau))}{\sigma_s \sigma_i}  \, \mathrm{d}t} 
\end{equation} 

where $\mu_{j}$ and $\sigma_{j}$ are the mean and standard deviations of the corresponding signals. This is basically the Pearson's correlation coefficient usually used in correlation analyses \cite{Pearson}, as a function of a signal lag $\tau$. The value of $\tau$ that maximizes the modulus of the correlation function  gives the error in time alignment of the signals $\Delta t$, while the value of $r$ gives a measure of the similarity of the signals, with 1 denoting perfect correlation and -1 perfect anti-correlation. 

\begin{figure}
\includegraphics[width=16.2cm,trim={1cm 0 0 0},clip]{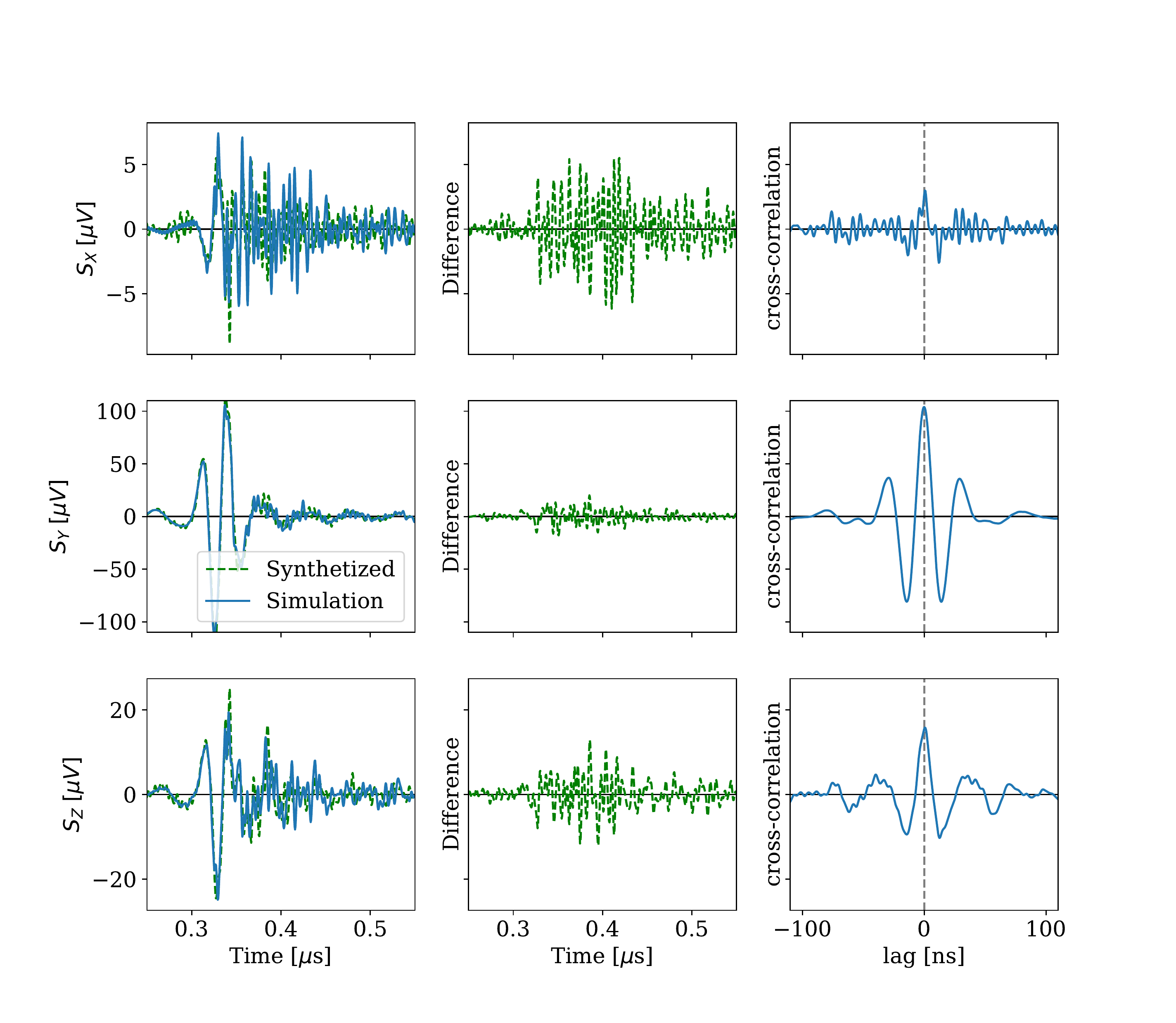}
\caption{ Example signals on the three polarization channels (left) for the selected antena of fig. \ref{Fig:ConicalStarshape}, showing the difference between the simulated and synthesized signal (middle) along with the correlation function r($\tau$)  multiplied by the signal maximum (right). Note that all the rows share the same scale for the ordinate. The maximum of the correlation is achieved close to lag 0, where the maximum correlation r=0.4 for $S_X$, 0.98 for $S_Y$ and 0.81 for $S_Z$  }
\label{Fig:CorrelationExample}
\end{figure}

An example of the evolution of the correlation function with $\tau$ for the 3 channels of the antenna selected in fig \ref{Fig:ConicalStarshape}, scaled with the signal amplitude, is presented in fig. \ref{Fig:CorrelationExample} (right column). In this particular case, the signal on the X channel is very small and incoherent, and although qualitatively the synthesized signal looks similar to the simulated one, the sample-to-sample difference is of the order of the signal itself. In this case r=0.4, and was very close to actually being classified as an anti-correlation with lag 12ns, looking at the negative maximum that corresponds to r=-0.34.  The signal in the Z channel is 4 times larger (but still of the order of the expected $\sigma_{noise}$) and the correlation improves significantly, with r=0.81. The Y channel is where most of the signal power is received, the sample-to-sample difference is of the order of 10\%, and r=0.98. In this case, the difference comes mostly from a high frequency component in the interpolated signal that is not present in the reference simulation.









\begin{figure}
  \includegraphics[width=15cm]{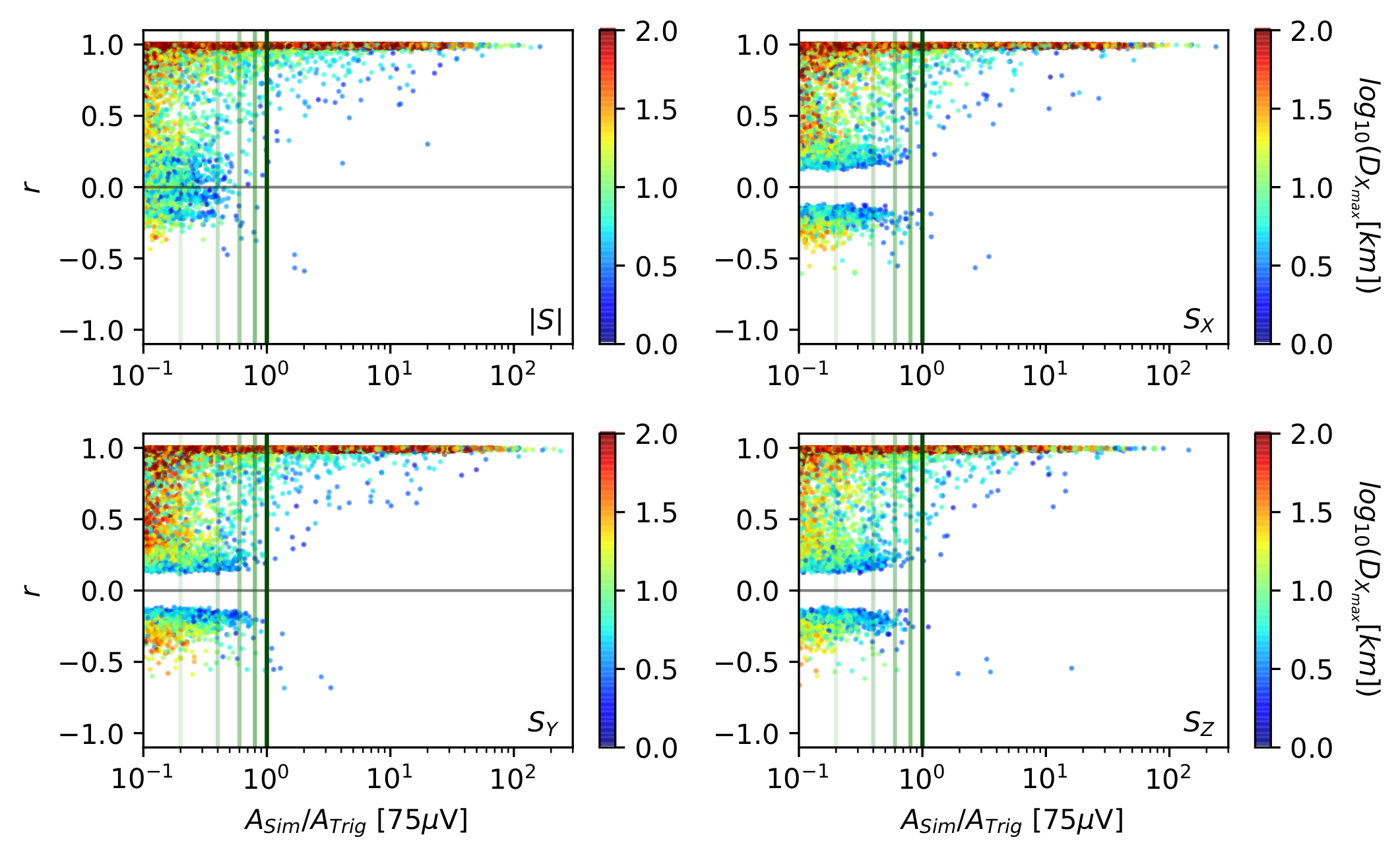}
  \caption{Correlation coefficient on each polarization and their weighted average (top-right, see text) for all the signals on the simulation library, as a function of the signal amplitude. Color indicates the distance from the antenna to ${X}_{max}$}
  \label{Fig:Correlation}
\end{figure}

The  maximum cross-correlation r of all the signals in the simulation library in each polarization component is shown in fig. \ref{Fig:Correlation}, as a function of the signal peak-to peak amplitude in units of the trigger amplitude, defined for this study to be 5 times $\sigma_{noise}$ (akin to a signal over noise). The lines corresponding to 1,2,3 and 4 $\sigma_{noise}$ are displayed to aid in the interpretation at other trigger levels. To give a measure of the overall agreement of the signal we also present in the top-left of this plot an average r, weighted by the amplitude on each channel, as a function of the maximum amplitude of the vectorial sum of the signals, defined as

\begin{equation}
|S|= \left(\sqrt{S_X(t)^2+S_Y(t)^2+S_Z(t)^2}\right)
\label{eq:vectorialsum}
\end{equation}

The figure shows that the interpolation performs exceptionally well once the signals are above 3-5 $\sigma_{noise}$ for antennas located farther than 3-5\,km away from $X_{max}$.

\begin{figure}
  \includegraphics[width=15cm]{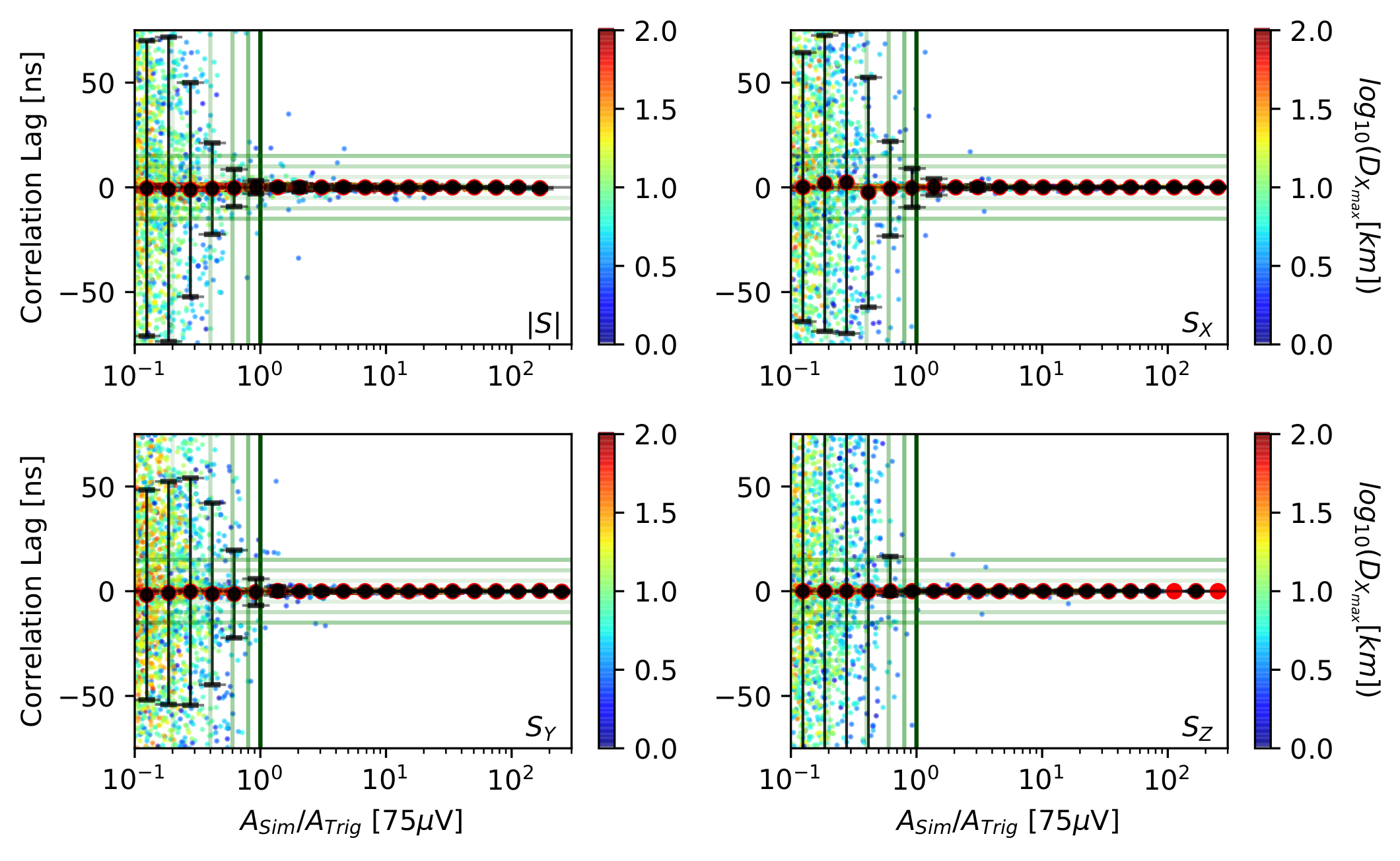}
  \caption{Correlation lag (with the weighted average on top-right, see text) for all signals on the simulation library, as a function of the signal amplitude for each polarization. Color indicates the distance from the antenna to $X_{max}$}
  \label{Fig:CorrelationTime}
\end{figure}

Looking at the values of $\tau$ that maximize the correlation, shown in fig. \ref{Fig:CorrelationTime}, we can see that once the threshold signal amplitude is reached, the timing of the synthesized signal is almost perfect, with most events requiring shifts of less than 1ns. This is a testament to the accuracy of the model proposed to align the timing of the interpolated signals.

The example shown in fig. \ref{Fig:CorrelationExample} illustrates that while a value of r close to 1 assures a good correlation, a low value of r is more difficult to interpret. For many applications, the signal synthesized for the X channel would be considered an acceptable rendition of what to expect at that antenna position, even if the relative sample-to-sample difference with the reference simulation is big. The peak-to-peak amplitude is within 8\% and the signal start time is accurate. For a larger signal, on the other hand, a value of r of 0.4 would be unacceptable. 

\subsection{Reduction in CPU time}

ZHAireS computes the electric field of the shower by adding the contribution from every advancement step (a track) of every electron and positron in the shower. For every track in the simulation, there is one call to the field calculation routine per antenna, so the total CPU time is proportional to the number of required antennas. Inclined showers cover more distance in the atmosphere before arriving to ground level, requiring more tracks to be computed. Consequently, the CPU time scales approximately as 1/cos($\theta$). 

The total CPU time required to compute the 16 test antennas of each event was on average 3.2 hr, varying from 1 hr at a 38 deg zenith angle to 10hr at 87 deg. In contrast, synthesizing the signal for 16 antennas with the method presented in this article takes on average 10 seconds, and its independent of the zenith angle. This is an improvement between 2.5 and 3.5 orders of magnitude in the required CPU time. To put it into perspective, making the microscopic simulation of the 55296 signals used for the comparisons took 1.3 years of CPU time and had to be run at a supercomputing center, while the synthesis of the same signals took only 7.5 hr of CPU time on a laptop. The interpolation method was implemented in python and was not optimized for speed, its performance can still be significantly improved. 

The gain comes of course at the expense of having to previously compute the 160 signals in the star-shape for each event, which took 13 years of CPU time. The gains from this method are capitalized only when the produced star-shapes simulations are to be re-used several times, as is the case in array optimization studies. 


\section{Performance of the method on signal observables}
\label{Sec:Performance}

We will focus our attention on the accuracy of the prediction for the peak-to-peak amplitude, which is generally used to evaluate the trigger probability when studying different array configurations. To asses the feasibility of using the method to characterize event reconstruction algorithms, we will also pay attention to the precision of the timing of the peak of the synthesized signals, used in the reconstruction of the incoming direction and position of $X_{max}$, and to the integrated power of the peak, used as a proxy for the reconstruction of the fluence, and thus the primary energy. 

 We have seen in the previous section that the interpolation has problems with signals with very low amplitudes and with antennas very close to the shower maximum, where the antennas are "embedded" in the shower and the signals lacks sufficient coherence. For this reason we set a quality cut removing antennas closer than 4\,km to the shower maximum for the remainder of the discussion.

\subsection{Maximum amplitude}
\label{Sec:Amplitude}

The peak-to-peak amplitude (from now on, the amplitude for simplicity) is regularly used to estimate the probability of detection, usually applying a simple signal-over-threshold trigger condition. If the amplitude exceeds the trigger condition, the antenna is considered to have detected the event. The condition can be applied on each polarization channel separately, or on a combination of channels to form an antenna-level trigger, as could be the vectorial sum of the signal defined in eq. \ref{eq:vectorialsum}. 

\begin{figure}
  \includegraphics[width=15cm]{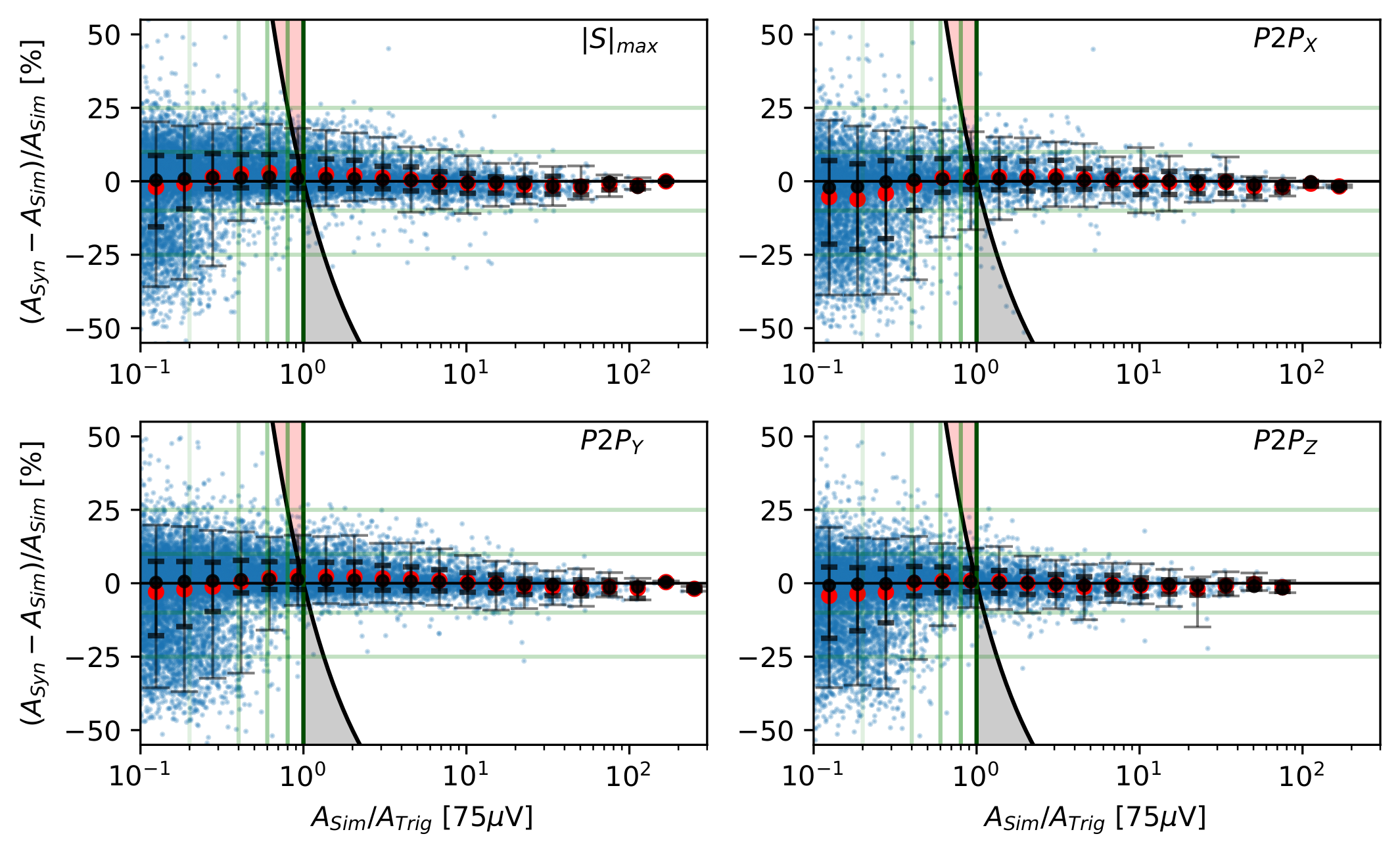}
  \caption{Relative errors in the amplitude of the synthesized antenna response, for the vectorial sum and for each antenna channel, as a function of the amplitude. Shaded areas indicate signals that would be miss-identified as false positives (red) or false negatives (gray). Vertical lines indicate 1, 2, 3, 4 and 5 (the trigger level) times the galactic noise. Horizontal lines indicate a 10 and 25\% relative error. Black dots indicate the median of the amplitude bin, while red dots indicate the average. Thick (thin) error bars contain 68.3\% (95\%) of the events around the median}
  \label{Fig:VoltageIErrors}
\end{figure}

The accuracy of our method to give the maximum of the signal is illustrated in fig. \ref{Fig:VoltageIErrors}, where the relative error in the amplitude on each channel is shown as a function of the amplitude. Like in fig \ref{Fig:CorrelationExample}, vertical green lines show the amplitudes corresponding 1 to 4 times $\sigma_{noise}$, to aid in the interpretation of the figure for different threshold values.

Since the distribution of the errors is not Gaussian, we use the mean and the median to characterize the accuracy of the method. To characterize the precision of the method, we use the interval containing 34.15 \% of the population at each side of the median, akin to one standard deviation, and the interval containing 47.8 \% of the population at each side of the median, akin to 2 standard deviations. We will call this precision estimates the 68\% and 95\% tolerance intervals (t.i.)

The close proximity between the mean and the median shows that the bulk of the events are symmetrically distributed around the median, and only the outliers are skewed. This is also evident from the increase in the asymmetry of the external error bars. Across all the channels, we see that the maximum amplitude is synthesized with an average accuracy well within 2\%, and a precision above threshold well within -4/+7.5\% at the 68\% t.i., and within -8/+15\% at the 95\% t.i., diminishing with increasing amplitude. Furthermore, almost no outliers are found above a 25\% relative error, which can be considered as an absolute upper limit.

Studies working with the signal amplitude face bigger systematic uncertainties from other sources, like the choice of the simulation program (20\% \cite{TimsReview}) or the experimental error from the calibration of the antennas (15-20 \%\, \cite{SchroderReview}). Its safe to say that using synthesized signals instead of complete simulations won't be the major source of uncertainties.

\subsection{Triggering errors}
\label{Sec:Triggering}

A finite precision in the determination of the amplitude will induce false positives and false negatives in the discrimination of the trigger condition. Events with an under-estimation of the amplitude, falling within the lower shaded area in Fig. \ref{Fig:VoltageIErrors} will give false-negatives, while events with an over estimation of the amplitude falling within the upper shaded area will give false-positives.

\begin{figure}
  \includegraphics[width=15cm]{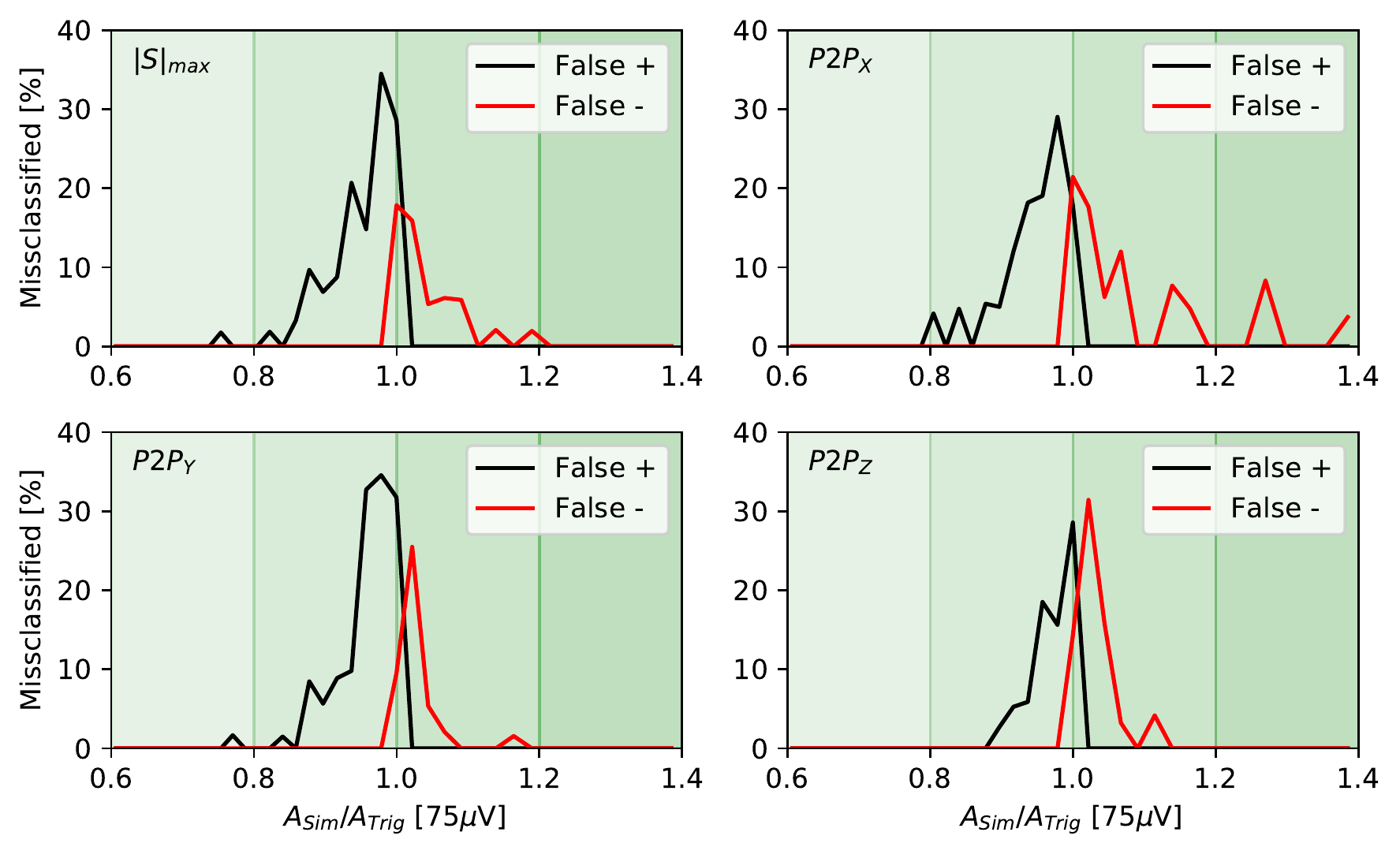}
  \caption{Miss-identification chance close to the threshold as a function of signal amplitude. Vertical lines indicate 4, 5 (the trigger condition) and 6 times the expected galactic noise. 
  }
  \label{Fig:VoltageChance}
\end{figure}

The chance of miss identification close to the threshold is shown in Fig. \ref{Fig:VoltageChance}, were it can be seen to quickly drop within 8\% of the trigger value, in accordance to the accuracy quoted in the determination of the amplitude. In our test library, less than 0.5\% of the signals where miss-identified as triggering and less than 0.25\% as non-triggering, regardless of the polarization channel. Note that when building event trigger conditions that involve multiple antennas, the miss-identification chance will be reduced even further.

\subsection{Integrated Power}
\label{Sec:Fluence}

The integral of the fluence of the electric field at the ground plane can be used to reconstruct the energy of the particle that originated the cascade  \cite{AugerFluence}. The fluence at each antenna location is calculated by integrating the absolute value of the Poynting vector in a time window of 200ns around the signal maximum. As a proxy to this quantity we use the integrated power received by the antenna, in the same time window.

\begin{figure}
  \includegraphics[width=15cm]{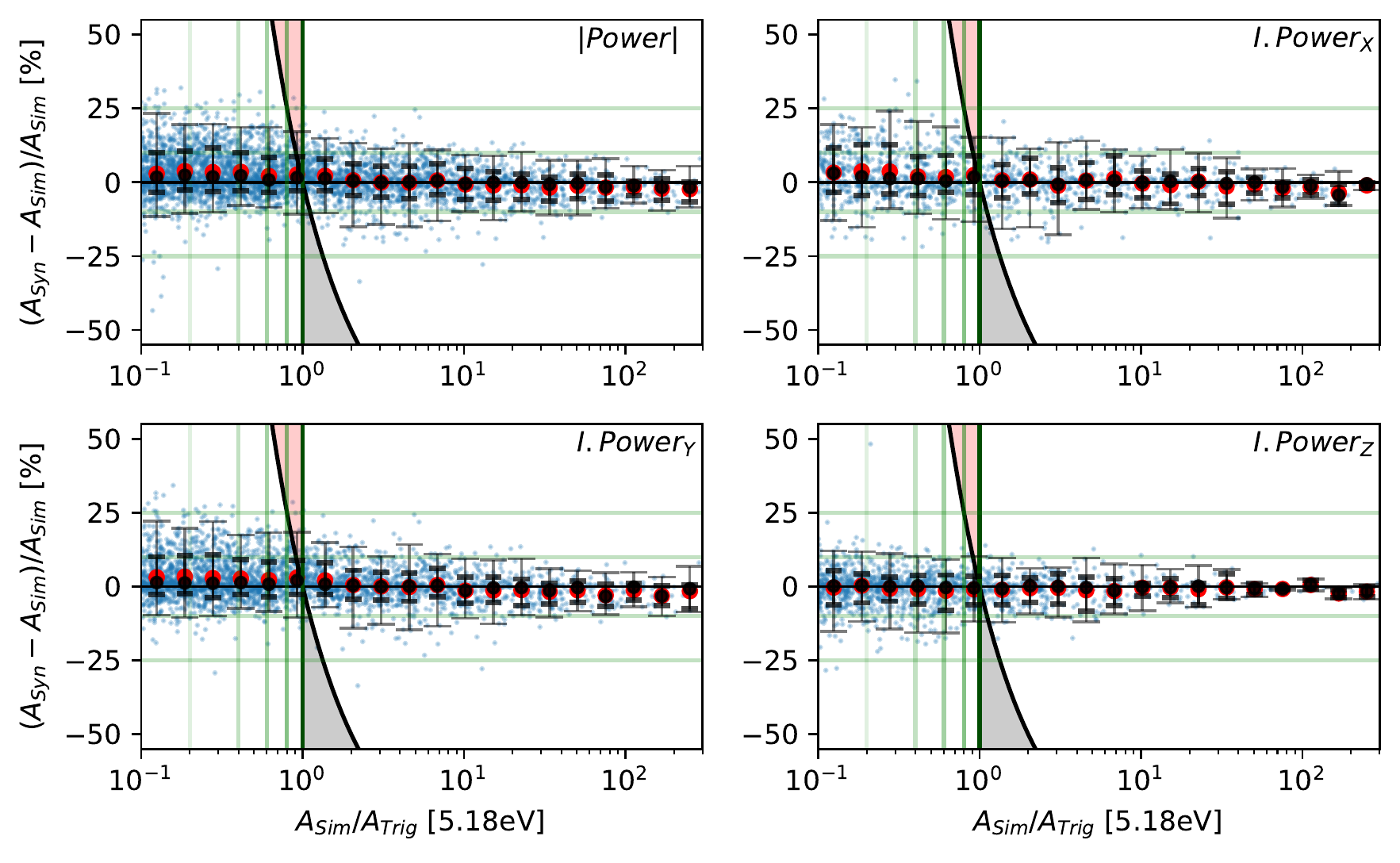}
  \caption{Relative errors in the integrated power of the synthesized antenna response for the Hilbert envelope of the complete signal and for the Peak-to-Peak amplitude in each polarization direction. Elements of the figure are described in fig \ref{Fig:VoltageIErrors}}
  \label{Fig:PowerIErrors}
\end{figure}

The error in the determination of the integrated power is shown in Fig. \ref{Fig:PowerIErrors}, following the considerations presented in section \ref{Sec:Amplitude}. The average synthesized integrated power has a precision and accuracy comparable to the one found for the signal amplitude, well below the expected systematic errors coming from the antenna calibration, the reconstruction methods and the signal simulation itself, showing that this method can be safely used for studying the energy reconstruction algorithms. There is a slight trend in the average difference, implying that the method over-estimates the integrated power at signal amplitudes close to the threshold by 2\% and underestimates it by the same amount at high integrated power, a trend that is also visible for the signal amplitude. We will get some insight on this by looking at the performance of the method as a function of $\alpha$.

\subsection{Linear interpolation effects}

Despite having an almost unbiased estimation of the amplitude when considering the the full simulations set, some dependence of the quality of the synthesis of the signal as a function of $\alpha$ is expected to exist, as we are using linear interpolation for functions that are clearly non-linear.

\begin{figure}
  \includegraphics[width=7.6cm,trim=0.1cm 0 0 0.2cm, clip]{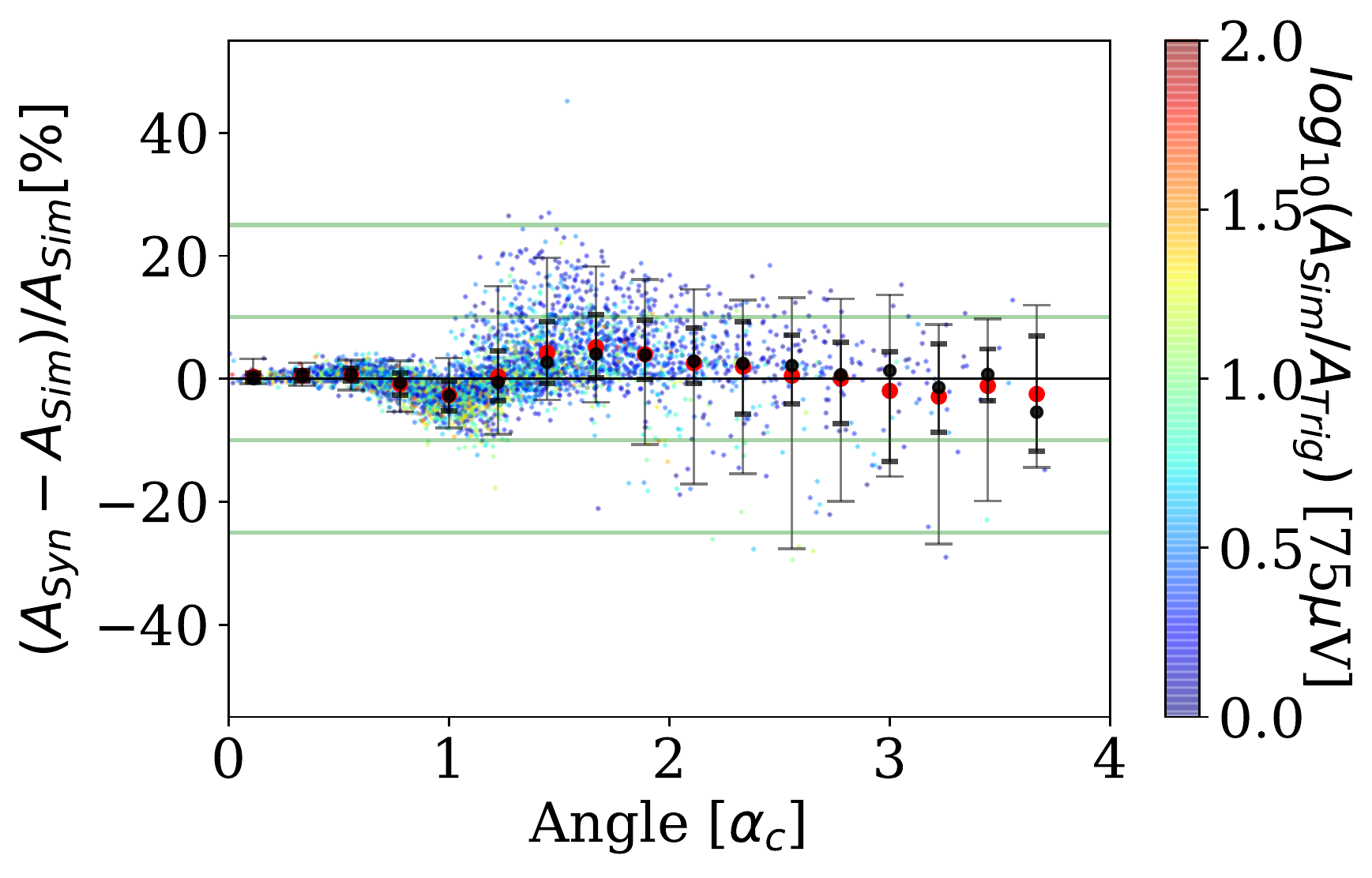}
  \includegraphics[width=7.5cm,trim=0.1cm 0.4cm 0 0cm, clip]{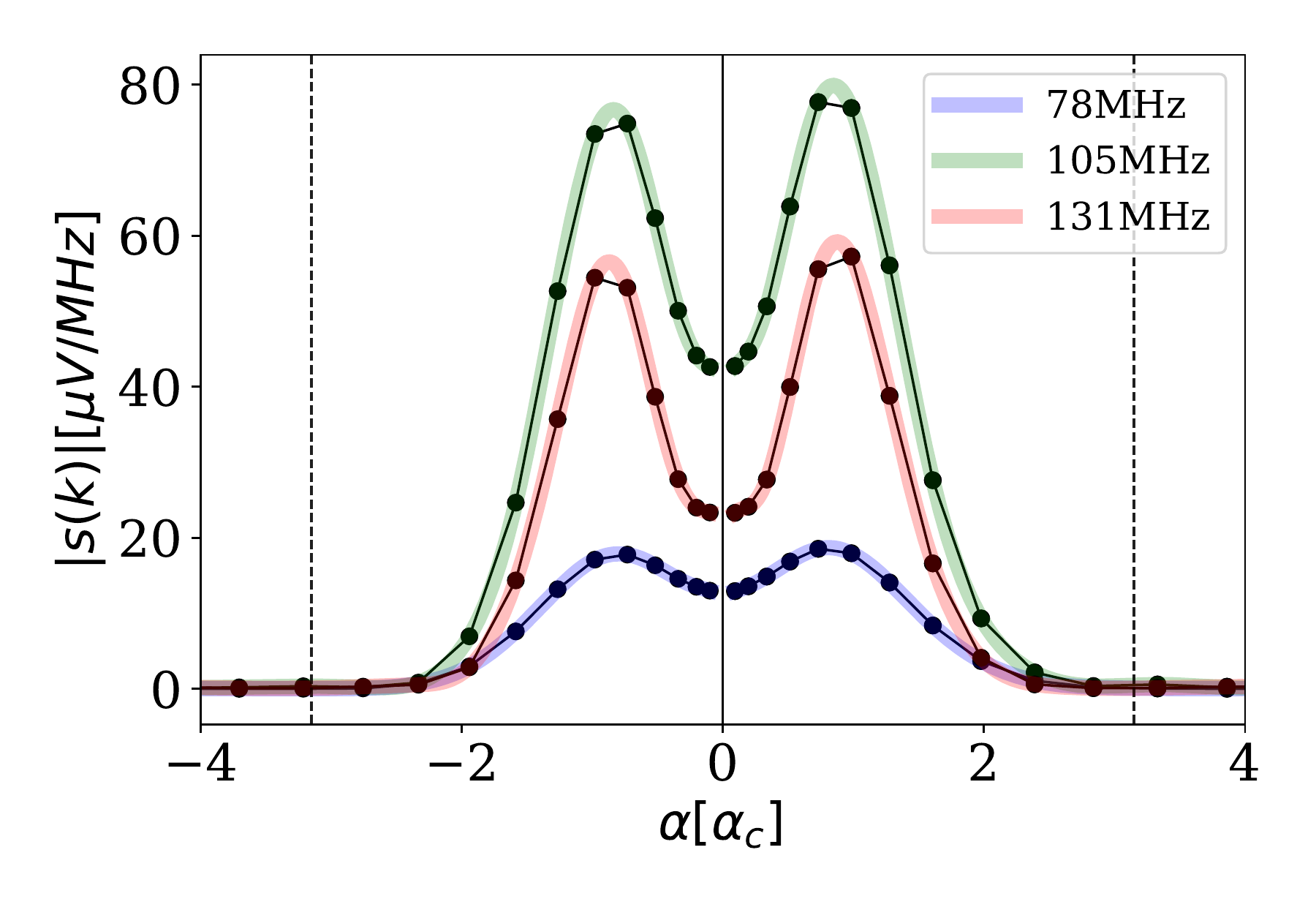}
\caption{Relative error of the peak-to-peak amplitude of the sum of the antenna voltage (left). Horizontal lines indicate absolute errors of 10 and 25\%. The amplitude of three different frequency components, and their linear interpolation are shown (right), to illustrates the angles at which the linear interpolation over/underestimates the amplitude. The vertical dashed lines indicate the angle after which the signals drop below the trigger threshold. See text for details.
  }
\label{Fig:AmplitudeAlpha}
\end{figure}

To illustrate how this affects the signal synthesis we show in fig \ref{Fig:AmplitudeAlpha}  (right), for our example event of fig. \ref{Fig:ConicalStarshape}, the lateral distribution function (LDF) of the amplitudes of the Fourier components of the Y channel at 3 different frequencies. The South arm of the star-shape is displayed at negative values of $\alpha$ and the North arm at positive values of $\alpha$. Looking closely at the linear interpolation between the antennas, we can see that when the curvature of the LDF is positive, a linear interpolation results in an overestimation of the signal, and when the curvature is negative in an underestimation. The errors are bigger at the places of maximum curvature, and the relative error is bigger at the inflection of the tail.

The $\alpha$ at which the curvature changes varies with frequency and the relative amplitudes of the frequency components of a signal vary with $\alpha$, making the final effect on the signal amplitude difficult to grasp. All frequency components will be underestimated close to $\alpha_c$, and all will be over estimated for $\alpha$ sufficiently big, with a different relative error for each frequency component.

Figure \ref{Fig:AmplitudeAlpha} shows the net effect on the vectorial sum of the antenna signals $|S|$. Individual components display the same behavior. The mean bias is of the order of -5\% at $\alpha_c$, negative as expected, and turns positive for bigger $\alpha$. Notice that outliers are caused by low amplitude signals, and that the precision is reduced with increasing $\alpha$. This also affects the integrated power, and is responsible for the trend observed in the average bias mentioned in the preceding section. Low amplitude signals, usually located at high $\alpha$, will tend to be over estimated while high amplitude signals, usually located at $\alpha_c$, will be under estimated.

These biases could be corrected with an ad-hoc parameterization, or reduced using a star-shape denser in $\alpha$. A non-linear interpolation, using the expected functional form of the LDF at each frequency, is a possible future optimization for the method.

\subsection{Peak Timing}
\label{Sec:Timing}

\begin{figure}
  \includegraphics[width=7.6cm,trim=0 0 0 0, clip]{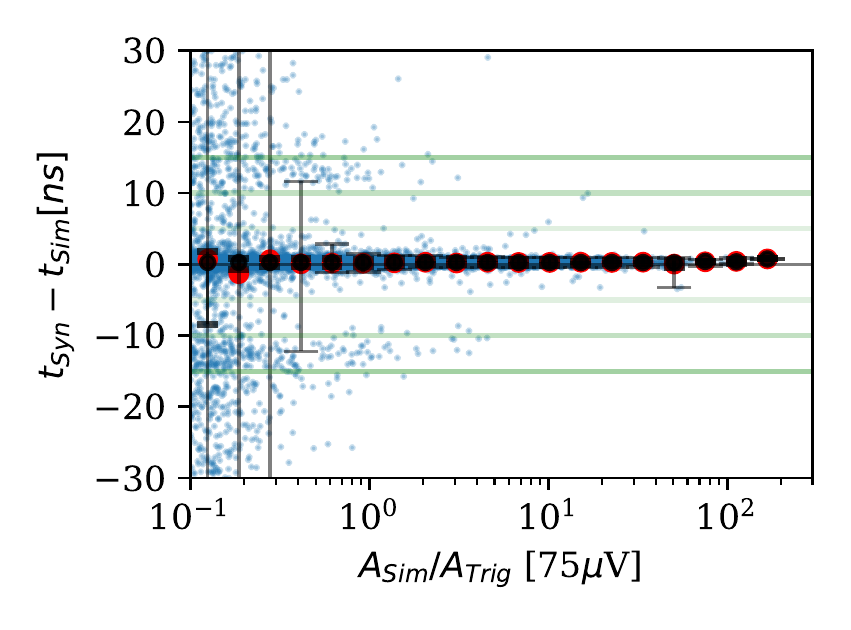}
  \includegraphics[width=7.6cm]{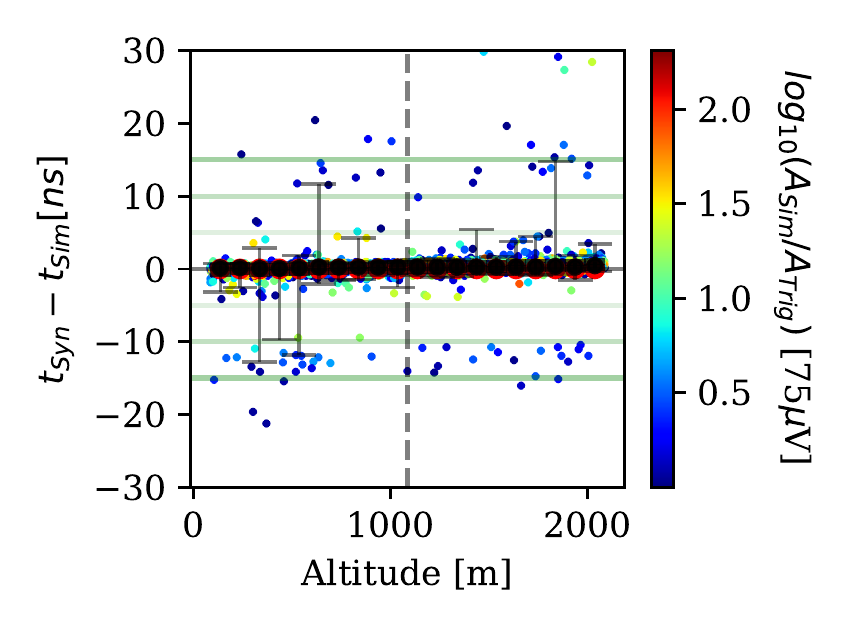}
  \caption{Hilbert envelope peak time error as a function of peak amplitude in the synthesized antenna voltage. Vertical lines indicate the 1, 2, 3, 4 and 5 (the trigger condition) times the expected galactic noise. Horizontal lines indicate absolute errors of 5, 10 and 15 ns.
  }
\label{Fig:TimeErrors}
\label{Fig:TopoTime}
\end{figure}

For the reconstruction of the shower incoming direction, and of the position of $X_{max}$, its necessary to have the correct time of arrival of the signal at the antenna. Since the signal will be always recorded in a noisy environment, the time of arrival is determined using the timing of the maximum of the Hilbert envelope of the signal. We have seen in section \ref{Sec:PerformanceCorrelation} that the lag between the synthesized and the simulated signal is close to 0, but differences in the shape of the signals could introduce shifts in the position of the maximum. Fig. \ref{Fig:TimeErrors} (left) shows the error in the timing of the maximum of the Hilbert envelope for the antenna response to the electric field. For signals below the trigger threshold it is sometimes difficult to identify a clear single peak structure in the signal, giving rise to outliers. A more robust algorithm than just looking for the maximum amplitude could be used to improve on this, if deemed necessary. Barring these outliers by considering the 95\% t.i., we can see that the timing accuracy is better than 1ns above 5 $\sigma_{noise}$, and is good to 2ns down to a more agressive 3 $\sigma_{noise}$ trigger threshold. This is comparable to the 2ns time stamping accuracy that can be achieved with current GPS technology \cite{GPStime}, indicating that signals synthesized with the method presented in this article can be used to develop and characterize reconstruction algorithms based on signal timing information without resulting in significant systematic errors.

\subsection{Local topography}
\label{Sec:TopoResults}

To test the performance of the method in dealing with different topographies, we generated an additional library of 1152 events in which the 16 test antennas were given an additional random vertical offset between -1000m and 1000m. For this library of events we used the same set of configurations as described in section \ref{Sec:Library}, but at a constant energy of $10^{18.6}$eV to maximize the statistics of signals above trigger.

\begin{figure}
  \includegraphics[width=15.2cm]{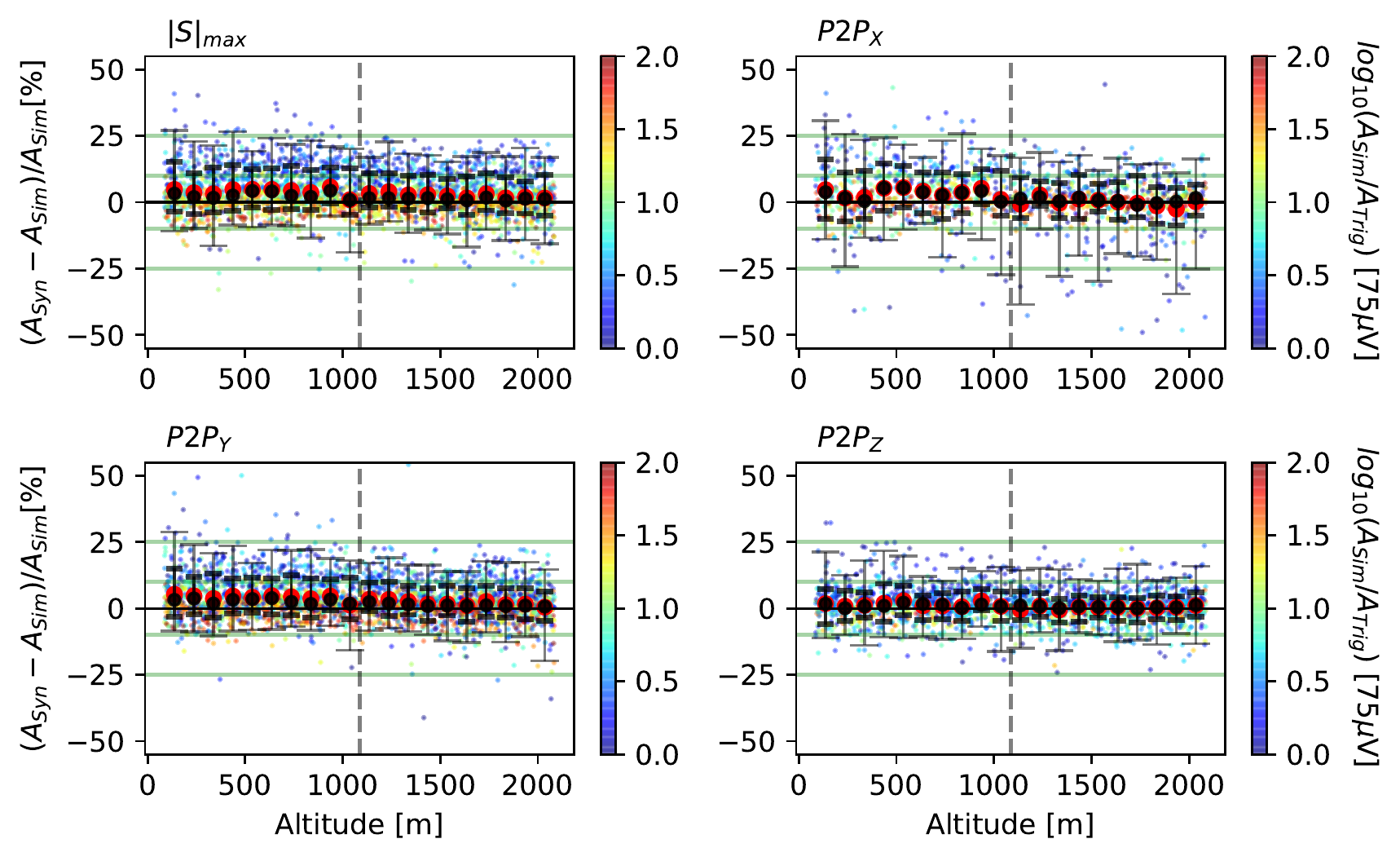}  
  \caption{Relative errors in the amplitude of the synthesized antenna response for the vectorial sum and for each antenna channel, as a function of the altitude of the antenna.  Horizontal lines indicate a 10\% and 25\% relative error.  The vertical line indicates the altitude of the ground plane.
  }
  \label{Fig:TopoAmplitude}
\end{figure}

Figure \ref{Fig:TopoAmplitude} shows that there is no bias and no change in the accuracy of the maximum amplitude of the synthesized signal associated to the topography correction introduced in section \ref{Sec:Method}. The timing accuracy is also not affected as seen in Fig. \ref{Fig:TopoTime} (right). Note that in this case the outliers correspond to the lower amplitude signals, which suffer from miss-identification of the true position of the signal peak.

\section{Conclusions}
\label{Sec:Conclusion}

We have developed a method for synthesizing signals with the precision of microscopic simulations but with a speed comparable to macroscopic methods, that opens the possibility to make accurate characterizations and optimizations of radio arrays for the detection of extensive air showers. 

The method has a precision on the estimation of the signal amplitude, usually used in triggering and reconstruction algorithms, of -4/+7\% at a 68\% t.i and of -8/+15\% at a 95\% t.i., well below other sources of systematic errors in the simulation chain. The method can handle differences in ground altitude related to the changes in local topography , without loosing its accuracy. The timing of the synthesized signals is accurate within to 2ns, enabling the use of the method for the study of geometry reconstruction algorithms. Since the full time trace of the signal is synthesized with good accuracy, the method could even be used for the production of datasets for the training of reconstruction algorithms based on neural networks, especially if noise is to be added later in the simulation chain. 

Improvements are possible, at the expense of increasing complexity or a narrowing scope of application. Fine tuning the antenna spacing to the expected position of the Cherenkov peak, and using a non linear interpolation of the amplitude are two clear lines of improvement. Since we know that the method starts having problems for low amplitude signals (below 3 $\sigma_{noise}$), setting a tighter maximum angle in the star-shape would already make the method more precise and efficient.

We focused the study on the performance for the response of the HorizonAntenna, but it would be equally applicable to other antennas or for the synthesis of the electric field filtered in the 30 to 300\,Mhz range. The method could also be used for synthesizing the unfiltered electric field, if the problems of the linear interpolation at the Cherenkov angle is addressed properly.

Once a sufficiently big library of events with the star-shape patterns is available, this method provides a 3 orders of magnitude reduction in the CPU time required to produce the simulations for any given study. Since computing time consumption is one of the biggest environmental impacts of building a radio array for the detection of extensive air showers, early adoption of this method could significantly reduce the $\rm{CO_2}$ footprint of such projects.

\section{Aknowledgments}

We would like to express our gratitude to Ewa Holt for providing us the algorithms to interpolate signals we started from and to Eric Hivon for his involvement in improving the phase interpolation. A special mention goes to Charles Timmermans for fruitful discussions and careful reading of the first draft, and to the GRAND group at IAP for their feedback throughout the development of this method.

This work was supported by the APACHE grant (ANR-16-CE31-0001) of the French Agence Nationale de la Recherche. This work has made use of the ExoAtmos cluster hosted by IAP. Most of the simulations were performed using the computing resources at the CC-IN2P3 Computing Centre (Lyon/Villeurbanne, France), partnership between CNRS/IN2P3 and CEA/DSM/Irfu. Simulations were stored on hardware acquired with grant UNLP PPID/X042.



\end{document}